\documentclass[epj]{svjour}
\usepackage{graphicx}

\begin{document}

\title{On the role of volatility in the evolution of social networks}

\author{Daniele De Martino\inst{1} \and Matteo Marsili\inst{2}}

\institute{International school for advanced studies SISSA, via Beirut 2-4, 34014 Trieste, Italy, Istituto nazionale di fisica nucleare, sezione di Trieste \and The Abdus Salam ICTP, Strada costiera 11, 34014, Trieste, Italy}

\date{Received: date / Revised version: date}

\abstract{
We study how the volatility, node- or link-based, affects the evolution of social networks in simple models. The model  describes the competition between
order -- promoted by the efforts of agents to coordinate -- and disorder induced by
volatility in the underlying social network.
We find that when volatility affects mostly the decay of links, the model exhibit a sharp transition between an ordered phase with a dense network and a disordered phase with a sparse network. When volatility is mostly node-based, instead, only the symmetric (disordered) phase exists
These two regimes are separated by a second order phase transition of unusual type, characterized by an order parameter critical exponent $\beta=0^+$.
We argue that node volatility has the same effect in a broader class of models, and provide numerical evidence in this direction.
}
 
\maketitle

\section{Introduction}
The competition between order and disorder is by no means restricted to physics.
Also economies and societies -- as systems of many interacting individuals -- organize themselves in different (macroscopic) states, with different degrees of order -- informally interpreted as coordination on social norms, compliance with laws or conventions \cite{PYoung}. Besides all its inherent complexity, one important element of additional richness is that the relation between the degree of order in a society and the cohesion of the underlying social network is not unidirectional as in physics, where the topology of interactions is fixed \cite{5}.
Rather the degree of order in a society influences in important ways the density and topology of interactions.
The interplay between network's dynamics and collective behavior is important in many phenomena, ranging from informal contacts in labour market \cite{Topa} and peer effects in promoting (anti-)social behaviors \cite{6} to inter-firm agreement for R$\&$D \cite{8}. The structure of the networks involved in these phenomena is dynamically shaped by incentives of agents (nodes), be they individuals or organizations, who establish bilateral interactions (links) when profitable.

In addition, this interplay typically takes place in a volatile environment.
That is, the favourable circumstances that led at same point to the formation of a particular link may later on deteriorate, causing the removal or rewiring of that link. This combination of factors raises a number of interesting issues in statistical physics, as the collective behavior -- of e.g. processes of ordering \cite{14}, opinion spreading, \cite{Voter} and reaction diffusion \cite{Blasius} -- may radically change when they are coupled to the dynamics of the network they are defined on.

We shall here focus on the stylized mathematical description of this generic phenomena given in Ref. \cite{14}: Here the feedback between nodes and networks dynamics, arises from assuming that the formation of a link requires some sort of similarity or proximity of the two parties. This captures different situation: For example, in cases where trust is essential in the establishment of new relationships
(e.g. in crime or trade networks), linking may be facilitated by common acquaintances or by the existence of a chain of acquaintances
joining the two parties. In other cases (e.g. in R$\&$D or scientific networks) a common language, methodology, or comparable level of technical
competence may be required for the link to be feasible or fruitful to both parties.

This class of models reveals a generic behavior characterized by a discontinuous transition from an uncoordinated state characterized by a sparse network, to a coordinated state on a dense network. As discussed in Ref. \cite{14}, this agrees with anecdotical evidence which can be summarized as follows:

(i)\emph{Sharp transitions.} Observation on the spread of social pathologies\cite{6}, the growth of research collaborations, both scientific\cite{7} and industrial\cite{8}, suggest that networks can shift from a sparse to connected state in  short time spans.

(ii)\emph{Resilience.} Once a transition to a highly connected network has taken place, this setup can survive even to a reversion to unfavorable
conditions, e.g. the thriving performance of Silicon Valley during the computer industy crisis of the 1980s\cite{9}, or the recent development of open-sourcesoftware, sustained against large odds, thanks to a dense web of collaboration and trust\cite{10}.

(iii)\emph{Equilibrium coexistence.} Under apparently similar environmental conditions, social networks can be found both in a dense or
sparse state. A good illustration is provided, e.g. by the dual experience of poor neighborhoods in large cities, where neither poverty nor
other socioeconomic conditions alone can explain wheter there is a degradation in a ghetto with rampant social problems\cite{6}.

This paper focuses on analyzing the effect of node volatility in these simple models. Indeed, the effect of volatility is limited to link removal in Ref. \cite{14}, but the turnover of agents (i.e. node removal and arrival) may be an important factor in many real systems. Our focus here will be mostly on the statistical phenomenon, than on its interpretation in socio-economic terms. Indeed we find that the introduction of node volatility brings in a qualitative change, which can be described as a continuous phase transition with unusual critical properties.
In order to show this, we concentrate on the simplest model for which a full analytic treatment is possible. In the concluding section, we argue that this qualitative change is expected in a wider class of model, and it can have much stronger effects.

\section{model}

Our model is a variant of one in the general class of\cite{14}. It reproduces in
a stylized manner the mechanisms of co-evolution in social networks
aforementioned, and shows some common elements of the observed phenomenology
(i.e. sharp transitions, resilience, equilibrium coexistence);
in particular, it also shows how the alternative assumptions of link or node based volatility have profound
effects on the dynamics of network formation.
The model describes $N$ agents sitting on the nodes of a network, each of which is characterized by a variable $\sigma_i$ which represents the social norm (convention or technological standard) adopted by agent $i$. There are $q$ possible social norms, i.e. $\sigma_i\in \{1,\ldots,q\}$. In terms of statistical physics, the model can be thought of as a $q$ state Potts model defined on a graph of $N$ nodes, that evolves in a coupled fashion to the dynamics of the system. The rules of the dynamics are the followings:

  \begin{itemize}
    \item each node attempts to establish a new link with a randomly chosen node at rate $\eta/2$. The link is established only if the two nodes have the same color
    \item links are destroyed at rate $1$
    \item all the links of a node are destroyed with rate
        $\alpha$
    \item the color of a node is updated with rate $\nu$ to the color of any of its neighbors, unless the node is isolated. In the latter case the nodes takes a random color.
   \end{itemize}

So, the parameter $\alpha$ interpolates between two kinds of volatility. For $\alpha=0$ volatility only affects links and for $\alpha\gg 1$ it mostly affects nodes. With respect to the parameters introduced in Ref. \cite{14}, we observe that the link decay rate has been set to $\lambda=1$ and that, for notation convenience, we scaled by a factor $2$ the link creation rate $\eta$. As observed in Ref. \cite{14}, the color update rule is effective only for isolated nodes, in the long run,
and in that case the color is drawn at random. Since only links between same type agents are created, after a transient all nodes are either isolated, or connected to nodes of the same color. Therefore the particular way in which the neighbor is chosen is immaterial. For example, both a majority rule (most frequent color) or a voter-type rule (random neighbor) would give the same dynamics.
The model can be generalized to a probabilistic update rule for the colors introducing a finite temperature $T$ (see \cite{14}). Results do not change considerably as long as $T$ is small enough, so we shall confine ourselves to the $T=0$ case.
\begin{figure}[h]
\includegraphics*[scale=0.5]{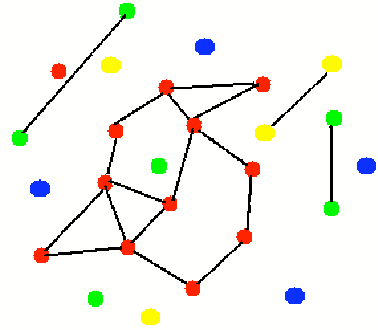}
\caption{Our model: links are formed with rate $\eta$ between nodes having the same color, links are destroyed with rate $1$,
all the links of a node desappear with rate $\alpha$ and  colors of isolated nodes are randomly updated with rate $\nu$.}
\end{figure}

Ref. \cite{14} has shown that for $\alpha=0$, the system shows an hysteretic transition in $\eta$ from a symmetric to an asymmetric state. The symmetric state is characterized by a sparse network, with average degree $\langle k\rangle<1$, with a symmetric distribution of colors. In the asymmetric state, instead, a dense network with  $\langle k\rangle>1$ arises, along with a dominant color, which is adopted by agents more frequently than the others. In this sense, the model shows  how order and disorder are intimately related with the dynamics of the social network in a volatile environment.

In what follows, we solve the model in the stationary state for $N\to\infty$, for all the values of $\alpha$. We find that the $\alpha=0$ behavior is generic for all $\alpha<1$, but the transition is softened as $\alpha$ increases. For $\alpha>1$ instead we show that the system is always in the symmetric phase. Hence, in terms of statistical mechanics, $\alpha=1$ is a second order critical point separating a phase with spontaneously broken symmetry from a symmetric phase.

\section{Theory}

If we call $n_{k,\sigma}$ the density of nodes with $k$ links and color $\sigma=1,\dots,q$ we have the following rate equations:
\begin{eqnarray}
\dot{n}_{k,\sigma} & = &
(k+1)n_{k+1,\sigma}-kn_{k,\sigma}-\alpha n_{k,\sigma}+ \nonumber\\
 & & {} + x_\sigma(n_{k-1,\sigma} - n_{k,\sigma}) \\
\dot{n}_{0,\sigma} & = &  \alpha(n_{\sigma}-n_{0,\sigma})+ n_{1,\sigma}-x_\sigma n_{k,\sigma}+ \nonumber\\
 & & {} + \frac{\nu}{q}\sum_{\sigma'=1}^{q}(n_{0,\sigma'}-n_{0,\sigma} )
\end{eqnarray}
where, for future convenience, we have introduced the dynamical variables
\begin{equation}
\label{ }
x_\sigma=\eta\sum_{k=0}^\infty n_{k,\sigma}.
\end{equation}
aking the sum over all $k$ of these equations and multiplying by $\eta$ we find
\begin{equation}
\label{dotnsig}
\dot x_\sigma=\frac{\eta\nu}{q}\sum_{\sigma'=1}^{q}(n_{0,\sigma'}-n_{0,\sigma} )
\end{equation}
which implies that, in the stationary state, each component has the same fraction $n_{0,\sigma}=n_0/q$ of disconnected ($k=0$) nodes. This is equivalent to a detailed balance condition for the density of the different components.

It is straightforward to derive an equations for the characteristic functions $\pi_\sigma(s)$ of the degree distribution $p_{\sigma}(k)=n_{k,\sigma}/\sum_q n_{q,\sigma}$ of the component $\sigma$. In the stationary state this reads:
\begin{equation}
(1-s)\frac{d\pi_{\sigma}}{ds}
=[\alpha+x_{\sigma}(1-s)]\pi_{\sigma}(s)-\alpha .
\end{equation}
The stationary solution is found by direct integration:
\begin{equation}
\label{charp}
\pi_\sigma(s)=\alpha\int_0^1\!dz z^{\alpha-1}e^{-x_\sigma(1-s)(1-z)}
\end{equation}
It is easy to see that this interpolates between a Poisson distribution, $\pi_\sigma(s)=e^{x_\sigma(s-1)/\lambda}$ for $\alpha\to 0$, which coincides with the result of Ref. \cite{14}, and an exponential distribution $\pi_\sigma(s)=\lambda/[\lambda+x_\sigma(s-1)]$ for $\alpha\to\infty$. Notice also that the average degree in component $\sigma$ is $\langle k\rangle_\sigma=\pi_\sigma'(1)= x_\sigma/(1+\alpha)$. This is precisely what one expects from balance of link creation and destruction of links in component $\sigma$.

Observing that $\pi_\sigma(0)=\eta\frac{n_{0,\sigma}}{x_\sigma}=\frac{\eta n_0}{q x_\sigma}$ we find an equation for $x_\sigma$ in the stationary state, which reads
\begin{equation}
\label{xsig}
G_\alpha(x_\sigma) \equiv \alpha x_\sigma\int_0^1\!du u^{\alpha-1}e^{x_\sigma(u-1)} = \frac{\eta n_0}{q}.
\end{equation}
Notice that the r.h.s. of Eq. (\ref{xsig}) is independent of $\sigma$. The variables $x_\sigma$ are determined by Eq. (\ref{xsig}) and the normalization condition, which takes the form
\begin{equation}
\label{norm}
\sum_{\sigma=1}^qx_\sigma =\eta.
\end{equation}
The properties of the solutions of Eqs. (\ref{xsig},\ref{norm}) depend on the behavior of the function $G_\alpha(x)$, which are discussed in the appendix,
and can be classified in symmetric and asymmetric solutions.

\subsection{$\alpha>1$: The symmetric solution}

For $\alpha>1$ the function $G_\alpha(x)$ is a monotone increasing function (see appendix), Hence
Eq. (\ref{xsig}) has a single solution and Eq. (\ref{norm}) implies that $x_\sigma=\eta/q$ for all components $\sigma$. Hence Eq. (\ref{xsig}) yields the total fraction of disconnected nodes
\[
n_0=\frac{q}{\eta}G_\alpha\left(\eta/q\right)
\]
as a function of the parameters $q,\alpha$ and $\eta$. We can analyze the stability of the symmetric solution recalling that  $n_{0,\sigma}=\eta G_\alpha(x_\sigma)$. Then Eq. (\ref{dotnsig}) becomes a dynamical equation for $x_\sigma$
\begin{equation}
\label{dotxsig}
\dot x_\sigma = \frac{\nu}{q}\sum_{\sigma'=1}^{q}\left[G_\alpha(x_{\sigma'})-G_\alpha(x_{\sigma} )\right].
\end{equation}
Linear stability of the symmetric solution is addressed by setting $x_\sigma=\eta/q+\epsilon_\sigma$, with $\sum_\sigma \epsilon_\sigma=0$. Then to linear order
\begin{equation}
\label{stabsymm}
\dot\epsilon_\sigma= \frac{\nu}{q}G_\alpha'(\eta/q)
\sum_{\sigma'=1}^{q}[\epsilon_{\sigma'}-\epsilon_\sigma]=
-\nu G_\alpha'(\eta/q) \epsilon_\sigma.
\end{equation}
Hence, as long as $G_\alpha(x)$ is an increasing function of $x$, the symmetric solution is stable. This is always the case for $\alpha>1$, as we shall see, it fails to hold for $\alpha<1$.

\subsection{$\alpha<1$: The asymmetric solution}

For $\alpha<1$ the symmetric solution still exists.
However the function $G_\alpha(x)$ now has a maximum for some $x_0(\alpha)$ (see the appendix) and $G_\alpha(x)\to \alpha$ from above as $x\to\infty$.
Therefore the symmetric solution becomes unstable when $\eta>\eta_+$ where
\begin{equation}
\label{ }
\eta_+\equiv q x_0(\alpha)
\end{equation}
because beyond that point $G_\alpha'(\eta/q)<0$.

\begin{figure}[h]
\includegraphics*[scale=0.85,angle=0]{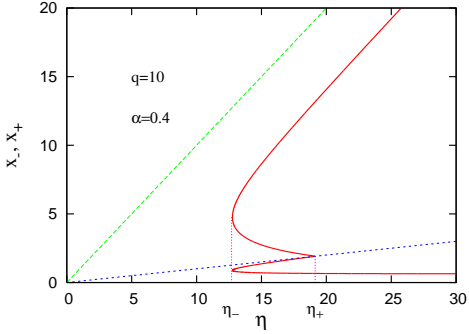}
\caption{Solutions $x_\pm$ as a function of $\eta$ for $q=10$ and $\alpha=0.4$. The dashed line $x=\eta/q$ separating the two curves is the symmetric solution.}\label{figxb}
\end{figure}

The occurrence of a maximum in $G_\alpha$ also implies that Eq. (\ref{xsig}) admits solutions with $x_\sigma =x_-<x_0(\alpha)$ for some $\sigma$'s and $x_\sigma=x_+>x_0(\alpha)$ for the other components. Since $x_\sigma$ is related to the density of a component $\sigma$, we shall call a component dense if $x_\sigma=x_+$ and diluted if $x_\sigma=x_-$.
As in Ref. \cite{14}, all solutions with more than one dense component are unstable. Indeed, by the same argument used to analyze the stability of the symmetric solution, a perturbation with $\epsilon_\sigma=0$ for all diluted components would grow as $\dot\epsilon_\sigma=-\nu G_\alpha'(x_+)\epsilon_\sigma$ on all dense components. These unstable modes correspond to density fluctuations across dense components and are clearly absent in the solution with one only dense component. These are the asymmetric solutions we shall focus on in what follows. There are $q$ of them, depending on which color is associated with the dense component. The variables  $x_\pm$ are determined by the system of equations
\begin{eqnarray}
G_\alpha(x_+) & = & G_\alpha(x_-) \\
x_++(q-1)x_- & = & \eta \label{norma}
\end{eqnarray}
This solution is shown in Fig. \ref{figxb}.
Actually, of the two asymmetric solutions the one with $x_+$ decreasing with $\eta$ is clearly unphysical as this would have a connected component with an average degree $\langle k\rangle_\sigma=x_+/(1+\alpha)$ which decreases with the rate $\eta$ with which links are formed. As in ref. \cite{14}, it is easy to see that only solutions with $x_+$ increasing in $\eta$ are  stable. Indeed, regarding $\eta$ and $x_-$ as functions of $x_+$ in Eq. (\ref{norma}) we find
\[
\frac{d\eta}{d x_+}=1+(q-1)\frac{d x_-}{d x_+}=\frac{G_\alpha'(x_-)+(q-1)G_\alpha'(x_+)}{G_\alpha'(x_-)}.
\]
Consider perturbations of the form $x_\sigma=x_++\epsilon$ for the dense component and $x_\sigma=x_--\epsilon/(q-1)$ for the others. Then by a derivation analogous to that leading to Eq. (\ref{stabsymm}), we find
\[
\dot\epsilon=-\frac{\nu}{q}\left[G_\alpha'(x_-)+(q-1)G_\alpha'(x_+)\right]\epsilon
=-\frac{\nu}{q}G_\alpha'(x_-)\frac{d\eta}{d x_+}\epsilon.
\]
Given that $G_\alpha'(x_-)>0$, this implies that on solutions with $x_+$ decreasing with $\eta$, the perturbation $\epsilon$ grows unboundedly.

\begin{figure}[hb!]
\includegraphics*[scale=0.7,angle=0]{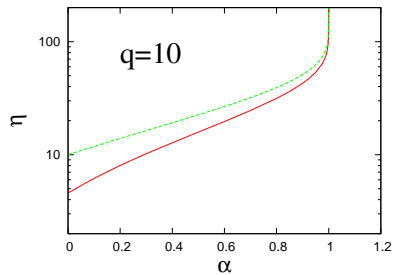}
\caption{Phase diagram for $q=10$. The symmetric phase extends below and to the right of the (full) line $\eta_-(\alpha)$ whereas above the (dashed) line $\eta_+(\alpha)$ only the asymmetric phase is stable. The coexistence region, where both phases are stable, is delimited by the two curves.}\label{figphasediag}\end{figure}

The asymmetric solution ceases to exist for $\eta<\eta_-$\footnote{We note, in passing, that the condition $d\eta/dx_+=0$ provides an equation which allows to determine $\eta_-$.}.
In the region $\eta\in [\eta_-,\eta_+]$ both the symmetric and the asymmetric solutions co-exist. The coexistence region, in the $\alpha,\eta$ plane is reported in Fig. \ref{figphasediag}.

\begin{figure}[hb!]
\includegraphics*[scale=0.2,angle=270]{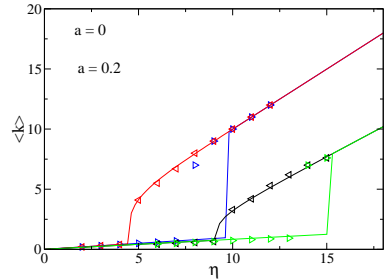}
\caption{Mean degree $<k>$ as a function of $\eta/\lambda$  for a system with $q=10$ colors, for $\alpha=0$ and $0.2$, simulations are
for systems of $1000$ nodes.}\label{figsweep}
\end{figure}

The practical relevance of the results derived so far is best discussed introducing an order parameter
\begin{equation}
\label{ }
m=\frac{x_+-x_-}{\eta}
\end{equation}
which is the difference in the density of the dense and diluted components. This
vanishes in the symmetric phase and is non-zero in the asymmetric one.
In Fig. \ref{figsweep} where we report the behavior of the average degree of the network
\begin{equation}
\label{ }
\langle k\rangle=\sum_{k,\sigma} n_{k,\sigma}  k=\frac{\eta}{1+\alpha}\frac{1+(q-1)m^{2}}{q}.
\end{equation}

Fig. \ref{figsweep} shows that as $\eta$ sweeps through the coexistence region the system undergoes an hysteresis loop: the degree jumps from low to high values at $\eta_+$ as $\eta$ is increased whereas when $\eta$ decreases from large values, the network collapses back to the symmetric phase when $\eta_-$ is crossed. In the case $\alpha=0$ \cite{14}, the symmetric phase is characterized by sparse networks, with a vanishing giant component. This is no more true when $0<\alpha<1$, specially close to $\eta_+$ \footnote{Indeed the condition for the presence of a giant component is $\langle k(k-1)\rangle_\sigma>\langle k\rangle_\sigma$ which, by a straightforward calculation, reads $\eta\ge q(1+\alpha/2)$. At the critical point $\eta_+=qx_0(\alpha)$ this reads $x_0(\alpha)\ge 1+\alpha/2$ which holds true for all $\alpha>0$.}. Numerical simulations fully confirm this picture, even though for finite systems the symmetric (asymmetric) phase is meta-stable close to $\eta_+$ ($\eta_-$) and therefore the transition occurs for lower (larger) values of $\eta$.

\subsection{The critical region: $\alpha\approx 1$}

\begin{figure}[h]
\includegraphics*[scale=0.25,angle=270]{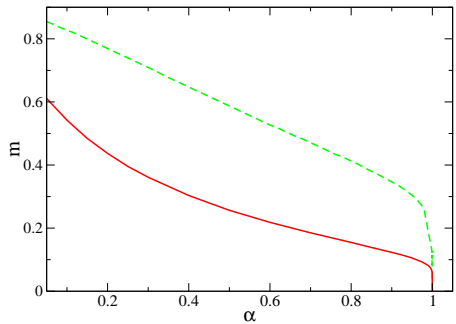}
\caption{Order parameter $m$ on the boundary of the coexistence region $\eta_-$ (full line) and $\eta_+$ (dashed line) as a function of $\alpha$ for $q=10$.}\label{figm}
\end{figure}

The behavior of the order parameter $m$ on the critical lines which confine the coexistence region is shown in Fig. \ref{figm}. This shows that the transition is continuous but with a peculiar critical behavior.
In order to shed light on this, the appendix shows that, asymptotically for $\alpha\simeq 1$
\begin{equation}
\eta_+(\alpha)=q x_0(\alpha)\simeq -q\log(1-\alpha)+c\log|\log(1-\alpha)|
\end{equation}
with $c>0$ a constant. A detailed asymptotic analysis of the limit $\alpha\to 1$ (see appendix) also shows that

\begin{equation}
m\sim 1/x_0(\alpha)\sim |\log(1-\alpha)|^{-1}.
\end{equation}
In terms of the usual description of critical phenomena, where $m\sim |1-\alpha|^\beta$, this model is consistent with an exponent $\beta=0^+$.
Indeed, the singular behavior of $m$ is very close to that of a first order phase transition.

\section{Conclusions}

The introduction of node volatility, in the simple model discussed here, makes the transition from a symmetric (disordered) diluted network to an asymmetric (ordered) dense network less sharp. Indeed when node volatility dominates, i.e. when the number of links lost per unit time by node decay outnumber those lost from link decay ($\alpha>1$), the transition disappears altogether, and the symmetric (disordered) state prevails. The phenomenology is strongly reminiscent of that of first order phase transitions (e.g. liquid-gas or paramagnet-ferromagnet) though the critical behavior is highly non-trivial.

The virtue of the particular model studied is that it allows a detailed analytic approach which allows one to gain insight on all aspects of its behavior. This model belongs to a general class of models which embody a generic feedback mechanism between the nodes and the network they are embedded in, which can be expressed in the following way: while the network promotes similarity or proximity between nodes, proximity or similarity enhances link formation.
This feedback allows the system to cope with environmental volatility, which acts removing links at a constant rate. Interestingly, the emergence of an ``ordered'' state plays a key role in this evolutionary struggle.

We believe the general findings discussed here will extend to the general class of models of Ref. \cite{14}. In particular, we expect the phase transition to be blurred by the effect of node volatility and to disappear when the latter exceed a particular threshold.

\begin{figure}[hb!]
\includegraphics*[scale=0.25,angle=270]{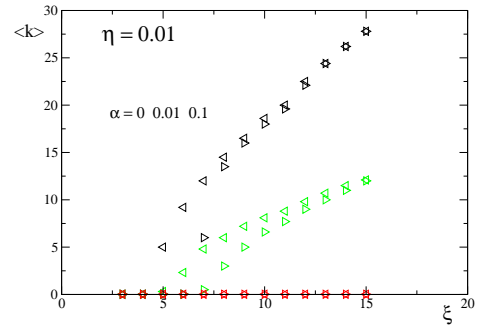}
\includegraphics*[scale=0.25,angle=270]{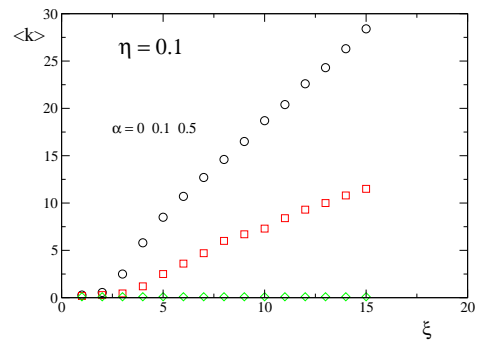}
\caption{Mean degree as a function of the rate $\xi$ of formation of links with neighbours of neighbours, for $N=1000$ ($\lambda=1$). Right: $\eta=0.01$, Left: $\eta=0.1$.}\label{vic}
\end{figure}

Actually, Fig. \ref{vic} shows that this is the case even for the model of Ref. \cite{13}. This is a model where link creation occurs either by long distance search at rate $\eta$ (as in the model discussed here) or through local search (on second neighbors) at rate $\xi$. Again links decay at unit rate. We refer the interested reader to Ref. \cite{13} for further details, for the present discussion let it suffice to say that the effects of (link) volatility are contrasted by the creation of a dense network with small-world features (a somewhat similar model with node volatility has been considered in Ref. \cite{12}). Fig. \ref{vic} shows that the effects of node volatility are very strong. Indeed, even a very small $\alpha$ reduces considerably the size of the coexistence region and the value $\alpha_c$ at which the latter disappears is also relatively small.

These results suggest that node volatility is indeed a relevant effect in the co-evolution of socio-economic networks, as it may affect in dramatic ways the ability of the system to reach a dense and/or coordinated state.

\appendix
\section*{Appendix}

The function $G_\alpha(x)$ can be written as
\[
G_\alpha(x)=\alpha\int_0^x\!dz\left(1-\frac{z}{x}\right)^{\alpha-1}e^{-z}
\]
For $\alpha>1$ we have
\[
\frac{dG_\alpha}{dx}=\frac{\alpha(\alpha-1)}{x^2}
\int_0^x\!dzz\left(1-\frac{z}{x}\right)^{\alpha-1}e^{-z}>0
\]

Notice also that
\[
\frac{d}{dx}\frac{G_\alpha(x)}{x}=-\alpha
\int_0^1\!duu\left(1-u\right)^{\alpha-1}e^{-ux}
\]
i.e. $n_0=G_\alpha(\eta/q)/(\eta/q)$ in the symmetric solution is a decreasing function of $\eta/q$. In addition $G_\alpha(x)\simeq x$ for $x\ll 1$, i.e. $n_0\to 1$.

For $\alpha=1-\epsilon$ we can approximate
\[
   G_{\alpha}(x)\simeq(1-\epsilon)\int_0^x du\left[1-\epsilon\log\left(1-\frac{u}{x}\right)\right]e^{-u}
\]
\[
     = (1-\epsilon)(1-e^{-x}+\epsilon(E_{i}(x)-\gamma))
\]
where $E_{i}(x)$ is the exponential integral function, and for $\epsilon \to 0$ we have $x_{\pm} \to \infty$ and $E_{i}(x)\simeq \frac{e^{x}}{x}$ so
\[
        G_{\alpha} \simeq (1-\epsilon)(1-e^{-x}+\epsilon/x)
\]
We have at the critical point $x_{+}+(q-1)x_{-}=qx_{0}$ where $x_{0}$ is such that $G'_{\alpha}(x_{0})=0$
so $\epsilon \frac{e^{x_{0}}}{x_{0}^{2}}=1$  and $x_{0}\simeq -\log\epsilon+2\log|\log\epsilon|$
by definition  $m=\frac{x_{0}-x_{-}}{x_{0}}$ and $x_{-}=x_{0}(1-m)$ $x_{+}=x_{0}(1+(q-1)m)$
so from $G_{\alpha}(x_{+})=G_{\alpha}(x_{-})$ we have
\[
  e^{-x_{+}}-\epsilon/x_{+}\simeq e^{-x_{-}}-\epsilon/x_{-}
\]
and then
\[
   \frac{qmx_{0}}{(1-m)(1+(q-1)m)}=e^{mx_{0}}(1-e^{-qmx_{0}})
\]
we have $mx_{0}\to c$ where $c$ is given by
\[
 c=e^{c}(1-e^{-qc})/q
\]


\begin{thebibliography}{13}
\bibitem{PYoung} H. P. Young 
{\em The Journal of Economic Perspectives}, {\bf 10}, 105-122, (Spring, 1996)

\bibitem{Topa} Topa, G., 
Rev. Ec. Studies {\bf 68}, 261 (2001).

\bibitem{5} M.Granovetter, Am. J. Sociol. {\bf 91}, 481 (1985)

\bibitem{Voter} C. Nardini, B. Kozma, A. Barrat, Rev. Lett. {\bf 100}, 158701 (2008); F. Vazquez, V.M. Eguiluz, M. San Miguel, Phys. Rev. Lett. {\bf 100}, 108702 (2008).

\bibitem{Blasius} T. Gross and B. Blasius, J. R. Soc. Interface {\bf 5}, 259 (2008).

\bibitem{6} J.Crane, Am. J. Sociol. {\bf 96}, 1226 (1991)

\bibitem{7} S.Goyal, M.J.van der Leij, J. L. Moraga-Gonzales, J. Polit. Econ. {\bf 114}, 403 (2006)

\bibitem{8} J.Hagedoorn, Research policy {\bf 31}, 477 (2002)

\bibitem{9} A.Saxenian, \emph{Regional Advantage: culture and competition in Silicon Valley and route 128} (Harvard university press, Cambridge,MA, 1994)

\bibitem{10} Y.Benkler, Yale Law Journal {\bf 112} 369 (2002)

\bibitem{12} J.Davidsen, H.Ebel and S.Bornholdt, Phy.Rev.Lett. {\bf 88}, 12, 128701 (2002)

\bibitem{13} M.Marsili, F.Vega-Redondo and F.Slanina, Proc.Natl.Acad.Sci. U.S.A. {\bf 101}, 1439 (2004)

\bibitem{14} G.Ehrardt, M.Marsili and F.Vega-Redondo, Phy.Rev. E {\bf 74}, 036106 (2006)




\end{thebibliography}
\end{document}